# Data analysis and visualization techniques for project tracking: Experiences with the ITLingo-Cloud Platform

**André Nobre Revés Barrocas**

**Alberto Rodrigues da Silva**

**João Paulo Saraiva**

INESC-ID, Instituto Superior Técnico, Universidade de Lisboa, Portugal

Considering the market's competitiveness and the complexity of organizations and projects, analyzing data is crucial to decision support on software development and project management processes. These practices are essential to increase performance, reduce costs and risks of failure, and guarantee the quality of results, keeping the work organized and controlled. ITLingo-Cloud is a multi-organization and multi-workspace collaborative platform to manage and analyze data that can support translating project performance knowledge into improved decision-making. This platform allows users to quickly set up their environment, manage workspaces and technical documentation, and analyze and observe statistics to aid both technical and business decisions. ITLingo-Cloud supports multiple technologies and languages, promotes data synchronization with templates and reusable libraries, as well as automation tasks, namely automatic data extraction, automatic validation, or document automation. The usability of ITLingo-Cloud was recently evaluated with two experiments and discussed with other related approaches.

**Keywords:** ITLingo-Cloud, Data Visualization, Data Analytics, Agile Project Management

## 1. Introduction

IT organizations are complex entities that intend to satisfy their customers' needs and adapt to market competitiveness, aiming to adapt to advancements in technology development. Project management (PM) practices have contributed to increasing the performance of organizations with better control of these processes. The lack of PM practices can bring huge losses to organizations, such as high costs, the product does not fit the needs, or the profit may not compensate the investment, compromising the quality of the product and customer satisfaction [1]. The failure of projects is often due to a lack of communication between stakeholders, poor planning, disregard for requirements engineering practices, and lack of technical documentation [2], [3].

Giving support to companies during the software development process is vital to reducing risks, uncertainties, and costs at different stages and to help meeting the deadlines. Hence, it is crucial to use software tools to support organizations' current activities and particularly support software development life cycles [4]. Since organizations generate data daily, data analysis and visualization techniques are essential to better understand actual and past project performance based on the analyzed data. Challenges such as handling conflicts and dependencies in resource allocation, planning under uncertainty, and facing threats during the execution of a project are frequent. Based on past project data, machine learning techniques can predict possible future undesired situations and recommend preventive actions [5]. These tools are crucial to help in planning work, maintaining the information available to the stakeholders to reduce inconsistencies through these processes [6]. In this scope, ITLingo is a research initiative that focuses on improving the rigor and consistency of technical documentation [3], promoting the productivity based on reusability of information and templates, and reducing inconsistencies and ambiguities. In the ITLingo project, some templates were developed to specify project plans, user requirements, and user interfaces aspects of software applications, among other business factors (e.g., PSL Excel Template, RSL Excel Template, ASL Excel Template) [3], [7], [8]. ITLingo-Cloud emerged inspired by this research focusing on overcoming the abovementioned situations supporting software development processes.

In this research, we propose an approach to support organizations and their projects, namely in what concerns the management and analyze information, providing mechanisms to guide software development processes. ITLingo-Cloud platform provides a collaborative allowing data synchronization with other technologies and approaches from the ITLingo initiative, namely PSL Excel templates [3]. This research involves analyzing and studying collaborative platforms and cloud tools to identify a solution to manage organizations and workspaces according to agile processes and support software development life cycles. We evaluate this approach by exploring the results with experiments and discuss the results produced with tests mainly considering the following qualities: learnability and usability. This paper presents and discusses the investigation's fundamental theoretical and technological concepts, including data analysis, visualization aspects, and project management theory. Furthermore we describe the solution proposed, presenting its architecture, implemented requirements, and technological aspects. Finally we evaluate and test the ITLingo-Cloud platform.

## 2. Background

We introduce and discuss general concepts underlying this research, namely on data analysis and agile project management aspects.

### 2.1. Data Analysis and Visualization

There are a large number of applications to whom data analysis is crucial. For example, in health domain, the capacity to present and analyze data understandably is critical to the success of public health surveillance. Also, in the business field, data mining is crucial to extract meaningful information from large amounts of information, helping organizations achieve their potential [9]. Statistical analysis techniques, namely linear and logistic regression, and artificial intelligence, are successfully applied for project performance prediction. Through the years, neural networks have been applied to areas such as pattern recognition, non-linear modeling, regression, and classification, in particular, to solve project scheduling problems, various operations planning and control activities, and forecasting project scheduling and control [10]. Furthermore, project management tasks have been tackled using neural networks, allowing to forecast of project completion for knowledge work projects. These technologies may guide managers to make complex decisions [11]. On the other hand, data visualization software allows to summarize and present information collected from different sources and assists developers and managers in visualizing performance in a shared way. The human brain responds better to visual information when compared to plain text [12]. Since organizations generate a massive amount of data, it is easier to analyze and understand it visually. Visual information improves communication, reduces misinterpretation, and clarifies massive or complex information, being essential to sharing ideas with accuracy and efficiency. Data visualization techniques use visual representation of data and usually perform data reduction, and transformation [12]. For instance, dashboards enable users to investigate and track trends, predict outcomes, and discover insights, using quick scanning and understating key metrics [13].

Common visualization techniques are [14] supported by the following charts: line charts, bar charts, scatter plots, and pie charts. The choice of the chart or visual representation depends on the data type that needs to be represented as well as the relationship among elements of the data, so it is vital to understand the concrete problem to choose the better representation to be used and, so, to show patterns, trends, and relationships correctly.

Descriptive statistics are used simply to describe and summarize the data logically, meaningfully, and efficiently, usually reported numerically in text, tables, or graphical forms, focusing on summarizing the characteristics of a data set (e.g., distribution, central tendency, variability). Inferential statistics involves using the available or representative sample data about a sample variable to make a valid inference or estimation about its corresponding unknown population parameter. This type of statistics uses a sample to make reasonable guesses about the larger population. (e.g., randomly selecting a sample of student graders to make estimates and test hypotheses about the whole population) [15].

### 2.1.1. Data Types

To report, analyze, and interpret data, as well as to understand and apply the findings, it is essential to have a basic understanding of data and variables. The combination of the data types that compose the dataset influences the choice of the data visualization to be implemented. At the highest level, data can be classified into two general categories: quantitative and qualitative.

**Quantitative data** is countable or measurable, while qualitative data is usually interpretation-based, descriptive, and not easily measured [15]. Quantitative data can be continuous and discrete (ratio) data. Continuous data are measured, have a constant sequence, or exist in a continuous range. This data type can be meaningfully divided into smaller or finer increments. Height, weight, temperature, and length are all examples of continuous data. Ratio data, considered another form of continuous data, have the same properties as interval data but the distinguishing property of ratio data is that it has a true definition of an absolute zero point. Age, height, weight, heart rate, and blood pressure are also Ratio data.

**Qualitative data** can be classified as nominal or ordinal. Nominal data (also called categorical data) represent types of data that may be divided into groups (e.g., race, sex). This data type can be classified as dichotomous (two categories) or polytomous (more than two categories). Ordinal data is data in which the values follow a natural order, while discrete data can be counted (e.g, age, educational level). Nominal data can only be classified, while ordinal data can be classified and ordered [15].

### 2.1.2. Data Visualization Techniques

Data visualization type depends on the specific datasets, needs, and the findings we want to show. Hence, it is crucial to understand the specific data and the most common data visualization techniques to transform the data into an effective visualization. Only a good choice of the chart type allows to properly represent patterns, trends, as well as relationships among elements of the datasets [14].

Common visualization techniques are the followings: Bar Chart, Line Chart, Pie Chart, Scatter Plot, and Histogram:

**Bar Chart.** Bar charts are one of the most common ways to visualize data and are used to compare quantities of different categories, revealing highs and lows at a glance and helping to understand trends in the data.

**Line Chart.** Line charts are one of the most frequently used chart types and can be used to compare changes and observe trends over a period of time. Line charts connect individual numeric data points being useful to visualize a sequence of values.

**Pie Chart.** Pie charts are helpful to compare the parts of a whole, showing relative proportions or percentages of information, being a fast way to understand proportional data. However, the pie wedges must be limited to six once it becomes too hard to meaningfully interpret the pie pieces when the number of wedges gets too high. Bar charts are a better option in cases that need more than six proportions to communicate.

**Scatter Plot.** Scatter plots are adequate to investigate relationships between different variables. This visualization technique provides an effective way to detect trends, concentrations, and outliers being helpful to guide where to focus some investigation efforts in the future. The story behind some data can be enhanced with a relevant shape or by adding a trend line.

**Histogram.** Histograms are adequate to analyze how the data are distributed across groups, helping to understand the distribution of the data, for example, the number of customers by company size or student grade on a specific exam. Histograms and bar charts look almost identical but display different types of data. Histograms present quantitative or numerical data, whereas bar charts depict categorical variables. The numerical data in a histogram is usually continuous [14].

Complementary relevant project-specific data visualization techniques in the scope of our research are Gantt Chart, Burn-up chart and Burn-down charts:

**Gantt Chart.** Gantt chart is a typical chart in project management tools, providing a complete overview of tasks, and dates, allowing project managers to drill down as needed to make informed decisions and hit deadlines for a successful project. This chart type is helpful to see what needs to be accomplished, representing the milestones, highlighting the start and finish dates elements for each deliverable, and illustrating resource planning to see how long it took project stakeholders to hit specific milestones and how that was distributed the tasks over the time, displaying the project schedule intuitively [14].

**Burn-down Chart.** A burn-down chart indicates how much work remains to be completed, keeping the end goal in mind and allowing to predict the likelihood of completing work during a period of time [16].

**Burn-up Chart.** A burn-up chart shows the project's progress over time and is an easy-to-understand figure of the status and rate of work done. This chart presents the increasing amount of work accomplished as a function of time that is reported on a regular basis [16].

## 2.2. Agile Project Management

Project management (PM) is a set of practices essential to maintain control of the project's scope, quality, schedule, budget, resources, and risk. These practices are crucial to compliance with the project objectives. PM is defined in the PMBOK as "the application of knowledge, skills, tools, and techniques to meet the project requirements", according to. PM practices involve identifying requirements and addressing the stakeholders' needs as the project is planned and carried out [2]. In this way, the performance increases consistently, improving customer satisfaction and speeding up the product's delivery according to the expectations [8]. Well-known international PM frameworks and guidelines are PMBOK Guide [17], ISO 21500 [18], IPMA ICB [19], and PM2 [20].

The process to be followed depends on the project's type, size, complexity, and duration [21].

Traditional approaches follow a "waterfall" model: The client communicates the plan and expects relatively clear results at the project's beginning [22]. In this approach, projects must define boundaries, and the plan should be followed as precisely as possible, finalizing within the planned time, budget, and scope. However, the reasons for the inappropriateness of these traditional approaches are the difficulty of dealing with the uncertainty and the change needs in current IT projects [23]. For instance, status values and principles such as the Manifesto for Agile Software Development were created to solve these restrictions [24]. This principle was written for agile software development but can also be applied to agile project management. Traditional approaches are more appropriate for projects with precise initial requirements and goals. On the other hand, agile project management approaches are better for projects with unclear or incomplete goals and where change is frequent [23].

Agile processes are focused on adaptability to changes during the project lifecycle and promote more communication and collaboration between stakeholders. In this approach, projects are managed iteratively, with frequent modifications in the project plan, focusing on fast implementation. Each interaction is usually short, achieving better control of uncertain projects. Agile project management also reduces risks and guarantees product quality [23]. Scrum and Kanban methods are popular agile methodologies used globally. Scrum defined few roles (Product Owner (PO), Scrum Master (SM), and the Development Team) with specific responsibilities. The typical size for sprints is one, two, or four weeks. The team manages the project with two artifacts: a Product Backlog, and a Sprint Backlog. To achieve the goals, typical activities such as Backlog Refinement, Sprint Planning, Daily Scrum Meetings, Sprint Reviews, and Sprint Retrospectives are performed. On the other hand, Kanban is not as prescriptive as Scrum. In Kanban, there are no defined roles and no emphasis on meetings or artifacts [25]. This methodology has its basis in the Just-in-Time (JIT) premise. It uses a visual Kanban board, limiting work in progress by reducing the number of tasks to be implemented [25].

## 3. ITLingo-Cloud System

ITLingo-Cloud is a multi-organization and multi-workspace collaborative platform to manage and analyze projects-related data, providing visual insights to aid in business decisions. Users can easily set up their environment and intuitively manage workspaces and technical documentation. This tool has been designed and developed considering the solutions available on the market, specifically to include standard features included in agile project management tools and inspired by emerging collaborative solutions like cloud IDE tools, which shall provide the necessary conditions to develop software quickly and with quality. ITLingo-Cloud shall support multiple technologies and languages studied in previous research projects (e.g., PSL Excel Template, RSL Excel Template), promoting data synchronization with predefined templates filled with project data. ITLingo-Cloud shall support automation tasks, automatic text extraction, or document automation, storing organizations' information and project data and using it to provide valuable dashboards. The system must provide integration with the ITOI system, an online integrated development environment (IDE), allowing to open and edit ITLingo-Cloud workspace artifacts stored in the database. Furthermore, this integration will enable software development and deployment in each workspace's scope. A user logged in ITLingo-Cloud will access and modify files directly, with the ITOI system. Besides, predictive data analytics models are being studied to monitor project costs, increase project productivity, and help managers make more informed decisions. The ITLingo-Cloud architecture encloses a relational database that stores these data and uses a popular Django framework.

## 3.1. Concepts and General Requirements

This section introduces the main components of the proposed solution, in particular, its supported users and roles, organizations, workspaces, agile processes, files management, data import, document automation, notifications, data analysis, and project prediction.

**Users and Roles.** Users may have different roles, determining the permissions to access the platform's pages, access certain information, and perform system actions that guarantee the access and management of the information. User roles may be changed easily at the platform, organization, and workspace levels, and the users' information may be searched, filtered, and sorted. User roles also can be chosen when some invitation is sent to be part of one specific organization or workspace.

**Organizations.** ITLingo-Cloud allows users to store, search and visualize several organizations' information. To better structure teams and facilitate collaborative work is possible to create several workspaces inside each organization space, named as "organization". Each organization may store files, namely libraries and templates (e.g., PSL Excel Template, RSL Excel Template), that may be used and filled with projects' data at the workspace level. The system stores and processes projects' data to provide intuitive statistics about organizations, intending to present and track their performance, such as: to analyze and compare budgets between project costs, to track projects to aid in business decisions. During the registration process, the user who registers in the system can create a new organization, thus becoming the organization manager of that organization. After the creation, the user has immediate access and control over it. A user can be added or invited to be a member of an organization that already exists in the platform. In this case, the organization manager of that organization needs to accept the access request to that organization. At any moment, a user with an organization manager role may access the list of the organization users from this organization, invite a new one or change user roles. Each organization provides statistics to analyze its data, namely on the analytics page and on its main page. It is possible to edit information related to the organization's name, organization activity type, or country on the settings page.

**Workspaces.** Workspaces are created in the scope of an organization. The main workspaces' properties are: project management process, status, project benefits, success criteria, general costs, and schedule information. These properties can be defined manually or automatically imported from files uploaded on the platform. Users can manage documents, the product backlog, and sprint backlogs. A workspace is a helpful space to support collaborative project management and contains several pages with several features. The system shall allow the storage of files (e.g., PSL Excel files) filled with project data. This data can be automatically imported from these files and consequently saving time. The users may select or merge the relevant data stored in these Excel files (e.g., related to the product backlog, sprint backlog items, project costs) and choose what they want to import into the database. This feature is particularly useful to promote better collaboration between teams or stakeholders since they may share dispersed files. ITLingo-Cloud shall provide visual insights such as statistics and intuitive dashboards to analyze workspace data, that may be imported or not. Users may generate project management reports by using the data stored in the database in the scope of each workspace.

**Agile Processes.** The product backlog, sprint backlog, and kanban board can be managed according to agile processes. It shall be allowed to store information associated with each product backlog item (or task), including files, images, and effort, among other relevant information. If the chosen process in (the workspace settings) is Scrum, users shall have access to the Sprint Backlog and Sprint History pages. The users shall manage sprints and respective tasks. The system shall allow to search the information, including sprint history and each sprint statistics page. If there is no initialized sprint, the system shall redirect to the sprint page creation. During the sprint creation, users shall specify relevant information, namely the sprint name, sprint schedule, and the stakeholders involved. Sprint information may be imported from Excel files or filled manually. The loading of data from the PSL Excel files detects if the task id already exists in the database or not to create or update its information on the database. If the tasks are created manually on the platform, it is necessary to provide all the information (e.g., name, type, priority, allocated people, task effort). ITLingo-Cloud shall allow the creation of tasks by importing items from the product backlog. With this functionality, the backlog item information is used to create a new task with these data, assigning a unique task ID. There are different types of effort defined for each task: (1) planned effort, (2) remaining effort, and (3) actual effort, which may be filled daily on the platform or imported from excel files. If the preferred software development process chosen is Kanban, users can choose this process in workspace settings. In this case, the Sprint Backlog page and Sprint History page are replaced by the Kanban page, which only displays the Kanban board with filters and the functionality to create new tasks. Kanban data also can be imported by Excel files, automatically generating the charts for these data.

**Files Management.** ITLingo-Cloud shall support the storage of multiple files, promoting the reusability and integration with ITLingo templates (e.g., PSL Excel Template). ITLingo libraries contain reusable specifications (e.g., defined in PSL or RSL languages), which can be used to create new specific specifications. ITLingo-Cloud shall support reusing these templates, importing data, and saving time by using these data in the scope of each organization and respective workspaces.

ITLingo-Cloud shall ensure that users can access the same documents' versions that stakeholders may share. ITLingo-Cloud shall allow selecting and merging data from these Excel files, allowing to select the content that the user may need (e.g., stakeholders, use cases, open issues, product backlog items).

**Data Import.** The stakeholders involved in each software project constantly generate data in organizations, including information related to requirements, bugs, budget, risks, and effort. When multiple stakeholders work simultaneously, there are problems with versions, data consistency, outdated information, and erroneous data. ITLingo-Cloud system shall allow import specific data from Excel files (namely based on PSL Excel templates) to speed processes, adding new information or updating existing data stored on the database. Furthermore, it shall be possible to partially select specific information from these Excel files and synchronize it with the information in the database. ITLingo-Cloud shall avoid duplicate information, detecting duplicated IDs of data elements stored in these Excel files. Hence, the ITLingo-Cloud shall update the information on the database if the ID already exists in the specific workspace, not duplicating data, otherwise will create the data on the database.

It shall be possible to select a specific PSL Excel worksheet(s) to be imported, namely: (1) all the project data (several worksheets) or only, (2) product backlog items, (3) sprint backlog or kanban tasks, and (4) project stakeholders.

**Document Automation.** Automatic report generation is an emerging technology that generates documents or reports containing text, tables, and figures about a specific topic [3]. ITLingo-Cloud shall provide a report automation system that shall use the data stored in the database to generate different reports. This feature is handy and saves time once the production of these technical artifacts can be a very monotonous and repetitive activity. Standard project management reports must follow certain practices or contain certain types of information. Some of them are identified in PMBOK [17]. Domingos Bragança carried out a study to understand what are the common reports in project management [26]. During his research, he designed the functionality to generate automatic reports in the PSL Excel template. The document templates collected and produced by Domingos Bragança were adapted to be used in ITLingo-Cloud.

The ITLingo-Cloud shall adopt the pre-defined templates stored on the platform and replace specific tags by organization and workspace data stored on the database.

**Notifications.** ITLingo-Cloud shall include a notification system to share relevant information and events that happen in the scope of organizations and projects. That notification system shall enable real-time notifications allowing directly read messages from the notification bar. Important information shall be automatically sent, namely invites to members of organizations or of specific workspaces. Also, accepting or rejecting the invite directly on the notification message shall be possible. Besides that, the system shall inform when one invite is accepted or rejected and when the project's planned end date is approaching. Invites to be member of organizations or workspaces are made even if a person does not have yet an account on the platform. In these cases, an email is sent to invite the user to register in the system. Depending on the invitation type, the invite to be member of a specific organization or workspace is made at the organization or workspace level. Supposing the user's email is already registered on the system, one invitation notification shall be sent, and it shall be possible to accept or reject the invite directly from the notification message.

**Data Analysis.** ITLingo-Cloud shall support the analysis of project-related data by summarizing and centralizing information. ITLingo-Cloud shall help to make more informed correct decisions with less uncertainty. The user shall explore the data through interaction with charts and obtain a summary of the characteristics and statistics of the organizations' and projects' data. ITLingo-Cloud supports dashboards at the organization and workspace levels to track progress, working hours, outcomes, and other relevant information. ITLingo-Cloud shall include (i) an organization main page that summarizes the organization's data and (ii) an organization analytics page that analyzes the data at a detailed level.

The data analysis mechanisms in the scope of each organization can be summarized as the following: (1) Compare organizational productivity changes over time, namely work completed over time, (2) track completed work per workspace, (3) analyze workspaces' status, (4) compare users involved per workspace, (5) analyze workspaces' costs, namely to better understand how to use the available budget, comparing planned cost and the current cost among workspaces. Figure 1 presents the ITLingo-Cloud organization analytics page.

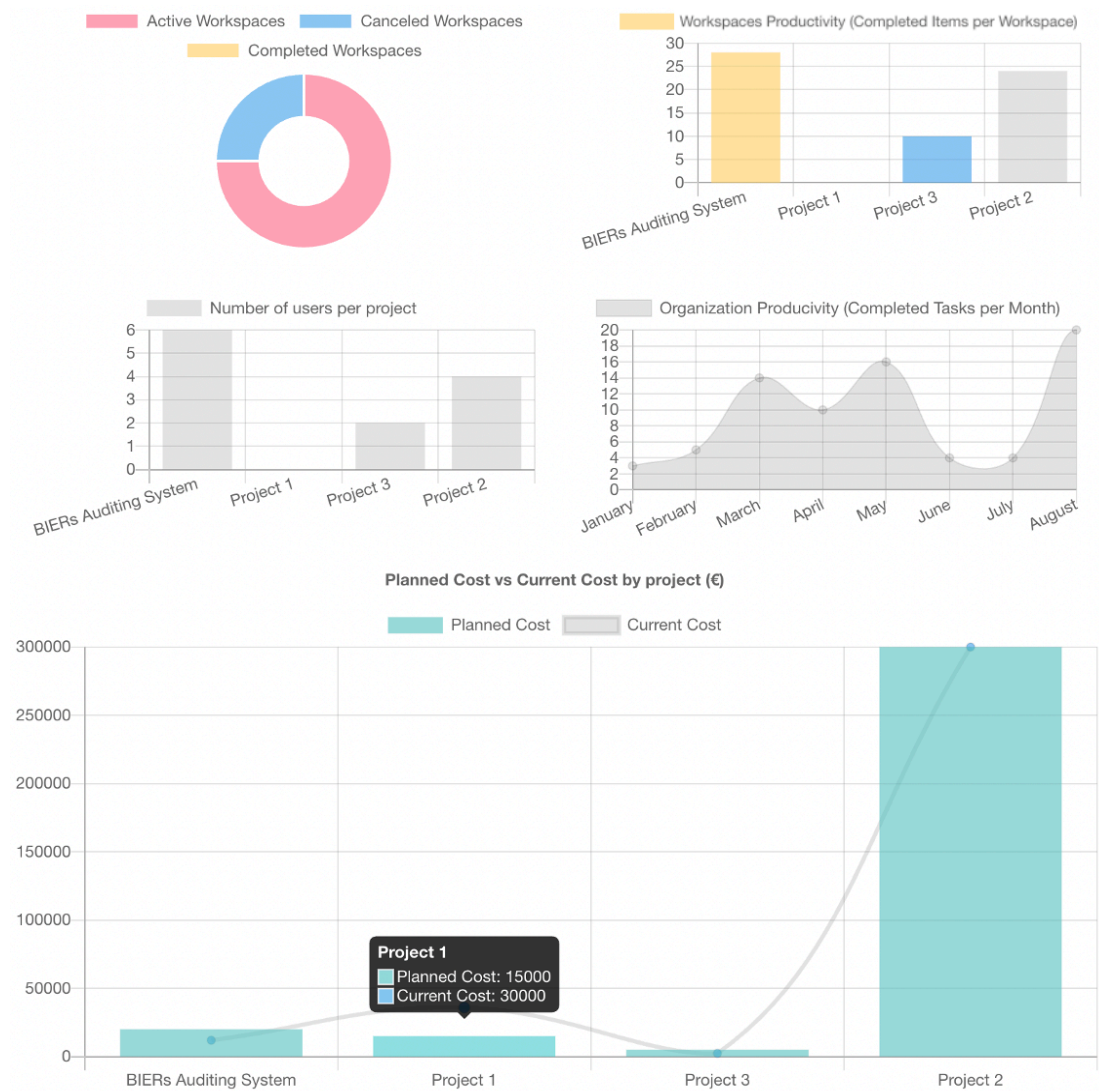

*Figure1 – Screenshot of the analytics page at the organization level*

At the workspace level, ITLingo-Cloud shall include (i) a workspace main page to summarize data among users and (ii) a workspace analytics page to show data at a detailed level. The data analysis mechanisms in the scope of each workspace can be summarized as the following: (1) Analyze the amount of the remaining work versus the time required to complete it in each sprint, (2) visualize the actual effort per day progressively in each sprint, (3) track the workspaces' productivity over time, illustrating the number of tasks in each sprint, and the actual number of work hours, (4) analyze product backlog items' status and types, (5) analyze the number of people involved in each sprint and individual performance, (6) analyze sprints' historical data, namely each sprint's open issues, the progression, the type of items solved, and tasks' remaining and actual effort per day.

Figure 2 presents the ITLingo-Cloud workspace analytics page.

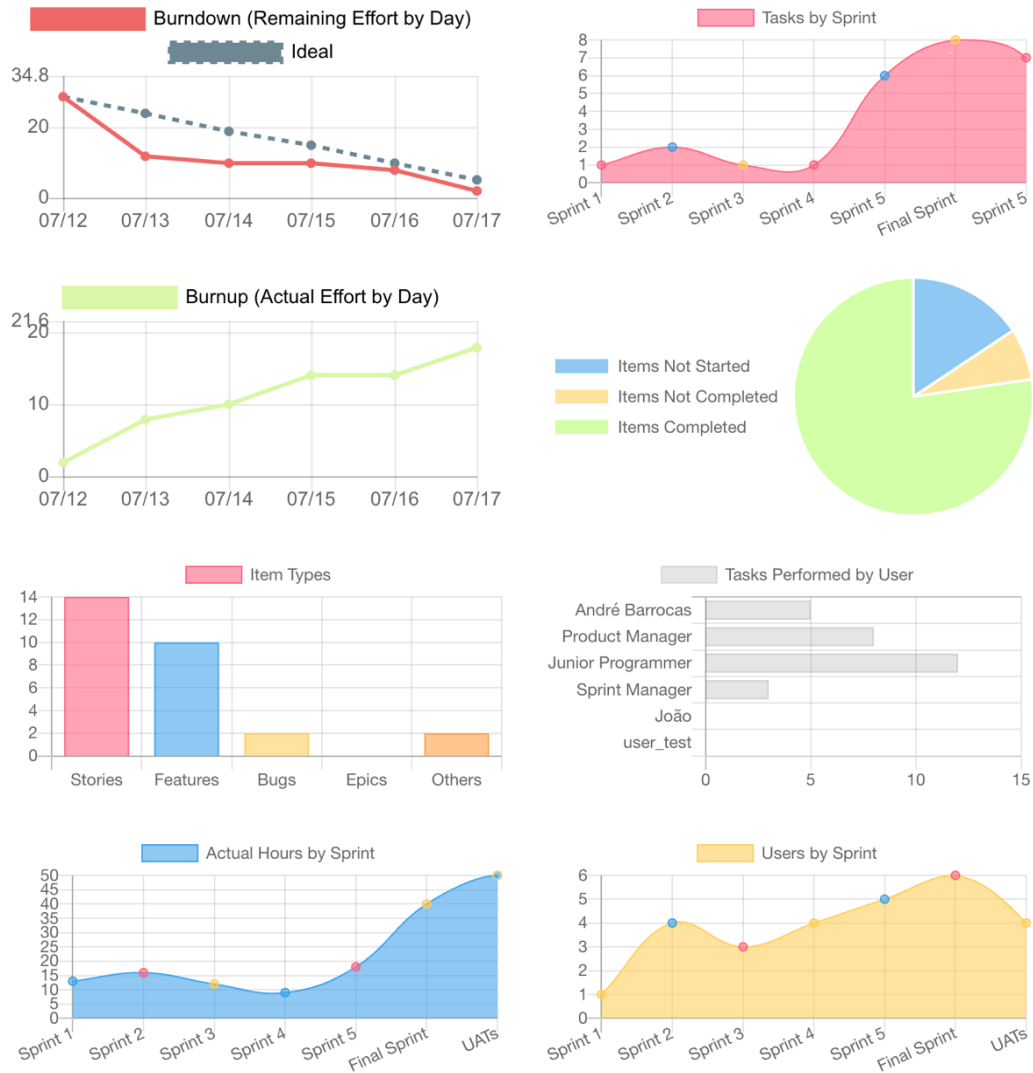

*Figure 2 – Screenshot of the analytics page at the workspace level*

**Project Prediction.** The system shall include predictive data analytics models to monitor project costs, to increase project productivity, and helping managers make more informed decisions. This models also shall be used to provide alerts when the project's end is approaching (e.g., a few days before the project's planned end date), the system could forecast planned dates based on data from past projects.

## 3.2. Architecture

The ITLingo-Cloud system has been implemented following a modular, iterative, and incremental approach. As illustrated in Figure 3, ITLingo-Cloud adopts a client-server-style architecture with three main elements: (i) a front end available to the client through a web browser, (ii) a back-end server, and (iii) a relational database.

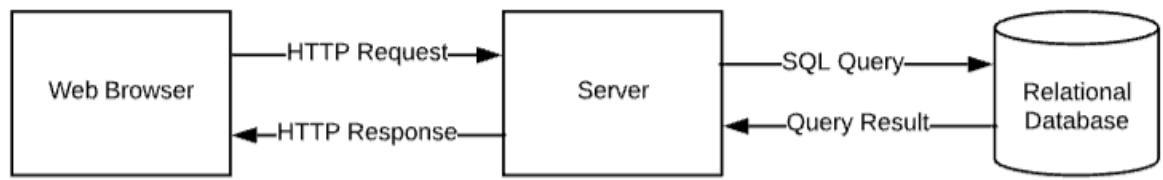

*Figure 3 – High-level Architecture of the ITLingo-Cloud system.*

The ITLingo-Cloud architecture encloses a relational database, in particular, implemented in the PostgreSQL database management system, that stores these data and uses a popular Django framework [27]. Users interact with the system through a web browser. The browser sends the HTTP request to the web server. The server translates the queries to SQL and sends them to the Database Management System (DBMS). The resulting query is returned to the Django application [27]; finally, the server returns an HTML page. In the front-end part we used Chart.js to develop the dashboards.

The web pages are responsive, so the ITLingo-Cloud can also be used on resized or smaller screens such as mobile devices (i.e. tablets or mobile phones).

## 4. Evaluation

We evaluate the ITLingo-Cloud platform through some experiments and a user session assessment with other researchers, teachers, and students, mainly to test collaborative and data analytics features of

ITLingo-Cloud. This activity is based on a fictitious organization named Bank of Investment and Economic Recovery (BIER) and a fictitious project discussed in the Information Systems Project Management course in 2022 [28].

### 4.1. User Assessment

We conducted a user assessment to evaluate the system and receive preliminary feedback from people not directly involved in this research. These tests were helpful to find problems and solve them, and evaluate the usability, some collaborative features, as well as the capacity to understand the data visualization mechanisms provided by ITLingo-Cloud, in particular, to test the ability of the users to analyze BIER's organization data and take some conclusions about it. In this evaluation process, we used usability tests with users in individual sessions to interact directly with the platform. Finally, we elaborated a questionary for the users to gather vital feedback to guarantee that the system meets the user's needs and analyze the obtained results.

The tests phase was performed between June and August of 2022. The questionary was answered by a group of 21 participants with ages ranging from 18 to 60 years old and at least a Bachelor of Science degree, namely 9 with a BSc, 9 with an MSc, and 3 with a Ph.D. degree.

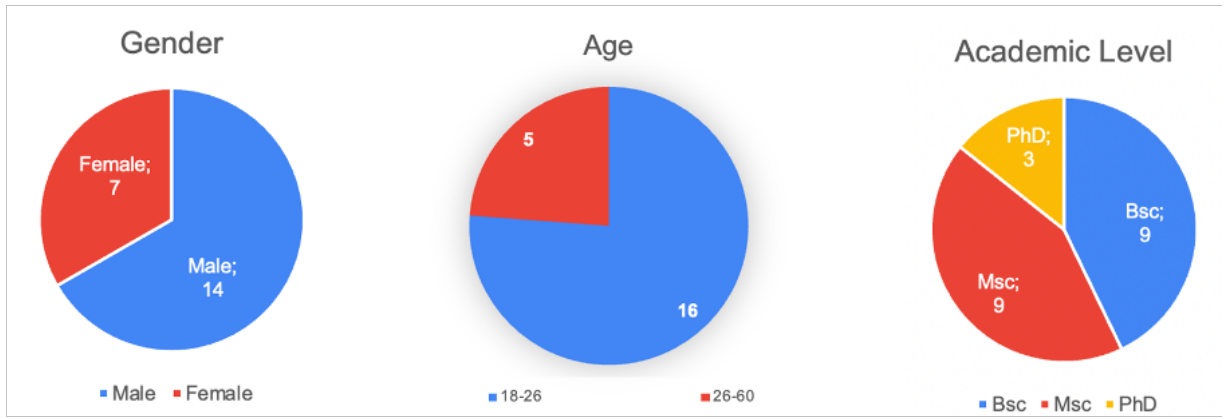

*Figure 4 - Demographic analysis of the group of participants in the user assessment*

Most participants had little professional experience, in particular, 13 participants with less than 1 year, 5 participants between 1 and 5 years, 1 participant between 5 and 10 years, and 2 participants with more than 10 years of experience.

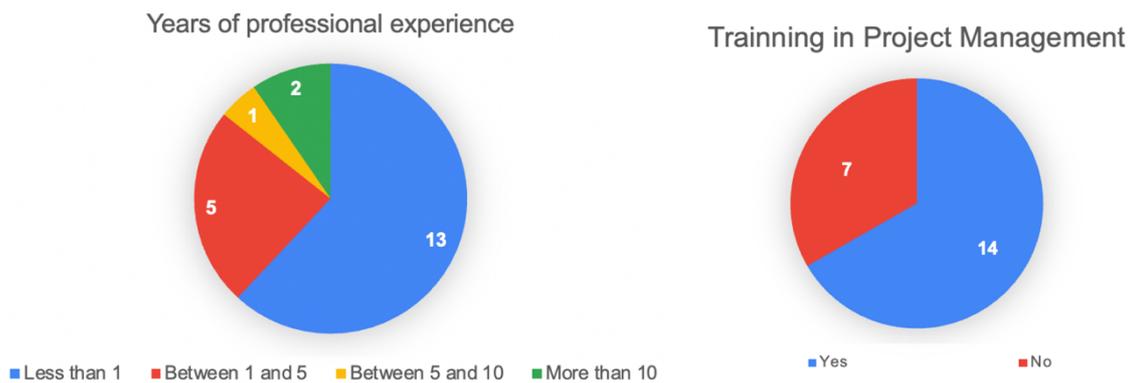

*Figure 5 – Background analysis of the group of participants in the user assessment*

By analyzing the users' background, it is possible to understand that most users have previous training in the project management field. In particular, 14 of the participants have previous training while 7 have no. By comparing the answers, it is possible to understand that most of the users have experience with project management tools while the minority have no experience with these tools.

The user pilot-user test session was conducted under the following conditions:
- The tests were conducted in a controlled environment without distractions (e.g., office or home environment).
- Realization of the tasks without previous use and learning of the system.

- The users were required to have an internet connection and a browser with good JS support.
- Users were free to think out loud and share ideas.
- The evaluator didn't interact with the users until the tests are finished (except in case of blocking errors).
- The session had a maximum of 45 minutes of duration.

To guide the user assessment test, participants received a script briefly describing the ITLingo-Cloud platform, explaining the goals of the activity, the BIERs' organization, and a fictitious project to be tracked with our system. After contextualizing the users, the steps to be followed to carry out the evaluation tasks were explained.

The user assessment session was divided into two parts: the first part involved testing some of the collaborative features of the platform, while the second involved the ability of users to use the data visualization features, in particular, to understand whether they were able to apprehend the preliminary information that the dashboards intend to transmit.

The first part included tasks 1 to 24 of the script, which involved the registration on the platform, creation of the organization and workspace, and management of the product backlog and sprints. The second part is referent to tasks 25 to 33 of the script. The users were asked to analyze the general statistics related to the BIER organization, its analytics page, the available workspace statistics, and its summarized analytics page, as well as the data analysis and visualization of a stored specific sprint in the historic.

The organization data used for test purposes were mainly imported from excel files, including information related to fictitious projects, to make it possible to compare data at an organization's global level, comparing information among its projects, stakeholders, productivity, effort, and performance.

In the end, participants were asked to fill in a questionnaire to rate the platform, suggest improvements, report errors, and answer the data analysis questions.

The questionary is divided into the following sessions:

**Section 1:** Respondent Characterization. The first three questions (Q1 to Q3) were focused on the general characterization of the participants with the following aspects: age, gender, and academic level.

**Section 2:** User Background Analysis. Four questions (Q4 to Q7) were directly related to user background analysis, namely related to the years of experience, previous experience in project management subjects, and related tools.

**Section 3:** User Experience Overall Assessment. Eight questions (Q8 to Q15) focused on test usability and learnability aspects of ITLingo-Cloud. We asked the percentage of completed the tasks proposed and asked participants to rate on a 5-Likert scale (i.e., from 1 to 5, 0-completely disagree, 5-completely agree) aspects related to the interface complexity, usability, feature integration, interface inconsistencies, and learnability.

**Section 4:** Data Analysis Questions. We asked five questions (Q16 to Q20) to evaluate the users' capacity to interpret the dashboards' information. There are a lot of possible aspects to ask related to the dashboards. Still, we focused on testing only two questions related to organization data analytics, two related to workspace data analytics, and one to interpret historic sprint data. We asked about organization productivity, costs, effort, hours of work, and completed items to check if the users understood the charts asking for a concrete answer.

**Section 5:** ITLingo-Cloud General Evaluation: This section has the highest number of questions (Q21 to Q30). In this part of the questionnaire, we asked about aspects related to the usefulness of specific features provided by the platform. We first asked participants to rate in a 5-Likert scale (i.e., from 1 to 5, 0—Do not know, 1-Very Low, 2 -Low, 3-Medium, 4-High, and 5-Very High) how useful the functionalities are to create and manage organizations, workspaces, import project-related data, manage the sprints and tasks, as well as the functionalities to analyze organization's and workspace's statistics, and sprint dashboards. Then, we include an open-ended question for users to provide additional comments (suggestions, problems, or bugs).

## 4.2. Results

As mentioned above, the questionary is divided into five sections. The first and second sections were already analyzed previously, describing participant-related aspects, namely the general characterization and background.

Section three analyzed ITLingo-Cloud usability and learnability aspects. Responses in this section revealed good interface results. When analyzing the answers, we concluded that 100% of the tasks were completed by the participants. For the remaining questions, we asked participants to rate on a 5-Likert scale (i.e., from 1 to 5, 1-completely disagree, 5-completely agree). Table 1 summarizes the average scores for these questions, based on which we may verify the following findings: Most participants would use the product again. When analyzing the results for the question Q11 - The system is more complex than necessary, we found that although most people disagreed with the question, there were still three people who answered 4, and one who answered 3. The vast majority of people agreed that the system is easy to use and easy to learn, feeling confident when using the product.

Some of the feedback provided by the participants on the open- ended question about the interface and usability aspects was: "The tool is very visual and intuitive, allowing a very interesting analysis of information." and "The platform is very complete and very visual, which I find great. Good job!".

*Table 1 – User experience overall assessment results*

| Questions | Average |
| --- | --- |
| I think I would use ITC again. | 4.43 |
| The ITC is more complex than necessary. | 1.81 |
| The ITC is easy to use. | 4.57 |
| The various features of ITC were well integrated. | 4.62 |
| ITC has inconsistencies. | 1.52 |
| I suppose most people would quickly learn to use ITC. | 4.71 |
| I felt very confident using ITC. | 4.67 |

(values on a 1–5 scale, 1-completely disagree, 5-completely agree)

Section 4 evaluates the users' capacity to understand the dashboards, as well as the ability to understand the information whose graphics are intended to convey and verify that the data analysis was done correctly, asking for concrete answers. Regarding the first question: Q16 - "Which project has more items solved in this organization?" all participants got it right. In the remaining questions, the vast majority got the answer right. The percentage of correct answers was 95.2% in the question Q17 - "Which project has the current cost higher than planned cost?", 95.2% in the question Q18 - "Which was the remaining effort foreseen in day 12?", 90% in the question Q19 - "Which sprint involved more hours of work?" and 90.5% in the question Q20 - "How many items were completed?".

The participants provided the following feedback on the open- ended question about the data analysis features: "Good monitoring and data analysis tool. You can monitor users and projects (active, canceled, and completed), among other tasks. One note would be to increase the size of the graphs to allow better visibility of the numbers on the scales of the graphs.". The last suggestion was followed, and the numbers of some of the graphs were increased as well as some colors were refined for a better understanding of the values.

Section 5 was very important to detect bugs and system improvements, particularly at the beginning of the testing phase. Users who tested the platform more intensively found a bug in the system that caused a message with no text when no description is provided. Another bug found was related to the drag-and-drop provided by the kanban board. This bug was related to the need to manually refresh the page for the kanban board items to be updated. These bugs were fixed at the beginning of the evaluation phase. Other improvement suggestions were: "I think the description should be an optional field (I ended up placing the name of the task in the description just to fill)" and "It would be interesting to add more integration features with other management tools, for example ". The first suggestion was followed, and task description is no longer a required field. The second suggestion was not implemented because this feature requires extensive future work. Nonetheless is an option to consider for future work.

In section 5, we also asked about aspects related to the usefulness of specific features provided by the platform. Participants rated, in a 5-Likert scale (i.e., from 1 to 5, 1-Very Low, 2 -Low, 3-Medium, 4-High, and 5-Very High), the usefulness of the functionalities of creating and managing organizations, workspaces, import project- related data, manage the sprints and tasks, as well as the functionalities to analyze organization's and workspace's statistics, and sprint dashboards. Table 2 summarizes the average scores for these supported features based on which we may verify the following findings: All answers were between 4 and 5 (4-High, and 5-Very High) in all questions, with the majority being 5. In question Q23 - How do you rate the usefulness of import product backlog data? there was one person who answered 3 (medium), and the remaining answers were between 4 and 5. With these results, we can conclude that users found all the features they evaluated useful.

*Table 2 – ITLingo-Cloud usefulness assessment results*

| Questions | Average |
| --- | --- |
| How do you rate the usefulness of creating and managing organizations? | 4.95 |
| How suitable is the platform for creating and managing workspaces? | 4.81 |
| How do you rate the usefulness of importing product backlog data? | 4.71 |
| How do you rate the usefulness of managing the sprints and the tasks? | 4.80 |
| How do you rate the usefulness of the sprint board and its filters? | 4.71 |
| How do you rate the usefulness of the organization's data analysis? | 4.90 |
| How do you rate the usefulness of the workspace's data analysis? | 4.95 |
| How do you rate the usefulness of the sprints' data analysis? | 4.95 |

(values on a 1–5 scale, 1-very low, 5-very high)

Regarding the open-ended questions of section 5, most of the participants did not answer. However, in addition to the feedback, improvements, and bugs that have already been mentioned before, those that answered provided encouraging comments and feedback, such as: "Very good Congratulations!", and "Congratulations on your work!", among the other positive feedback mentioned previously. The analysis of these comments led to the conclusion that this tool is useful and that the work was carried out successfully. To sum up, the results collected and analyzed in all sections had very positive scores. Usability experts like Nielsen and Landauer observed that a group of 5 testers is enough to uncover over 80% of the usability problems [29]. Since our questionnaire focuses on the usability and general evaluation of the ITLingo-Cloud platform, we may conclude that 21 participants are a fair number for an exploratory assessment, allowing us to identify significant flaws in the usability of such proposals. Furthermore, this assessment was also handy for detecting and solving a few bugs.

## 5. Conclusion

This research designed and developed the ITLingo-Cloud system, a collaborative platform that supports multiple software development processes and project management activities, allowing to manage and analyze organizations' data. ITLingo-Cloud is a platform that provides an appealing and easy-to-use interface that software developers, managers, and other stakeholders can use to keep a large amount of project-related information always accessible. With this work, we address the following concerns: (1) increase the performance of organizations; (2) managing projects more intuitively, allowing people to synchronize information and work collaboratively; (3) generating and managing project-related data, keeping it accessible to the major stakeholders; (4) improve collaboration among stakeholders within organizations and within workspaces; and (5) provide visual mechanisms to directly translate that knowledge on project performance into decision making support. Thus, ITLingo-Cloud platform supports these issues by providing an environment with reusability and adaptability features, supporting synchronization with other tools and technologies. This solution was evaluated through experiments and a user session assessment with several researchers, teachers, and students, mainly to test its collaborative and data analytics features.

This research identifies yet several aspects that may be addressed in future work.

First, the ITC may improve its data analysis capabilities to better monitor project costs through predictive data analytics models to increase project productivity, helping managers make more informed decisions. Furthermore, although the ITC already provides alerts when the project's end is approaching (e.g., a few days before the project's planned end date), the system could forecast planned dates based on data from past projects.

Second, the ITC support for multiple processes could be extended and improved. For example, the support of the Kanban process already implemented in the platform could be improved to be able to define customizable lanes and WIP limits.

Third, this research may also involve the completion of the document automation module to include more reports on the platform.

Finally, the integration of ITC with ITOI may still involve relevant research, namely including more common cloud IDEs features such as more developer tools, allowing software development and project management in an integrated platform.